\documentclass{article}
\usepackage{color}
\newcommand{\editR}[1]{{\color{black}{#1}}}
\usepackage[dvipsnames]{xcolor}
\usepackage[utf8]{inputenc}
\usepackage{wrapfig}
\usepackage[]{graphicx,xcolor}
\usepackage{tabularx}
\usepackage{booktabs}
\usepackage{textcomp}
\usepackage{amsmath}
\usepackage{amssymb}
\usepackage[]{graphics}
\usepackage{amssymb}
\usepackage{amsmath}
\usepackage{color}
\usepackage{ifpdf}
\usepackage{lipsum}
\usepackage{siunitx}
\usepackage{braket}
\usepackage{soul}
\usepackage{tikz}
\usepackage{appendix}
\usepackage{authblk}
\usetikzlibrary{automata,positioning}
\usepackage{dcolumn}% Align table columns on decimal point
\usepackage{bm}
\usepackage[margin=0.8in]{geometry}
\graphicspath{ {./pics/} }% пакет для задания полей страницы командой \geometry
\author[1]{D.~A.~Gromyko}
\author[1]{S.~A.~Dyakov}
\author[1,2,3]{S.~G.~Tikhodeev}
\author[1]{N.~A.~Gippius}
\affil[1]{Skolkovo Institute of Science and Technology}
\affil[2]{Faculty of Physics, Lomonosov Moscow State University}
\affil[3]{A.~M.~Prokhorov General Physics Institute, RAS}
\date{}                     %% if you don't need date to appear
\begin{document}

% \begin{frontmatter}
\title{Resonant mode coupling approximation for calculation of optical spectra of photonic crystal slabs \\Part II.}

% \author[inst1,inst2]{D.~A.~Gromyko}
% \author[inst1]{S.~A.~Dyakov}
% \author[inst2,inst4]{S.~G.~Tikhodeev}
% \author[inst1]{N.~A.~Gippius}

% \affil[inst1]{organization={Skolkovo Institute of Science and Technology},%Department and Organization
%             addressline={Nobel Street 3}, 
%             city={Moscow},
%             postcode={143025}, 
%             country={Russia}}
% \affil[inst2]{organization={Faculty of Physics, Lomonosov Moscow State University},%Department and Organization
%             addressline={Leninskie Gory, d.1, str.2}, 
%             city={Moscow},
%             postcode={119991}, 
%             country={Russia}}
% \affil[inst4]{organization={A.~M.~Prokhorov General Physics Institute, RAS},%Department and Organization
%             addressline={Vavilova 38}, 
%             city={Moscow},
%             postcode={119991}, 
%             country={Russia}}
\maketitle
    \begin{abstract}
    We propose further development of the resonant mode coupling approximation for the calculation of optical spectra of stacked periodic nanostructures in terms of the scattering matrix. We previously showed that given the resonant input and output vectors as well as background scattering matrices of two subsystems, one can easily calculate those for the combined system comprising two subsystems. It allows us to write a resonant approximation for the combined system and speed up calculation significantly for typical calculation problems. The main drawback of this approach is that the background matrix in such approximation was considered as constant which is not always sufficient if the energy range of interest is relatively wide. 
    %With the main source of method's inaccuracy being the error of the subsystems' background scattering matrices approximation, t
    The aim of this article is to solve this problem by utilizing more complicated approximations for the background matrices. In particular, we show that consideration of energy-dependent correction terms for the background matrices remarkably reduces the resonant energies' calculation error. Here we first consider a linear approximation, and although it is not suitable for large energy ranges, it is used as a base for a piecewise-linear approximation which allows one to keep the approximation error negligibly small with only a few sample points. Moreover, interpolation of the background matrices allows one to apply resonant mode coupling approximation in almost arbitrary large energy ranges. We also consider approximation of background matrices by an arbitrary matrix function and propose a technique to derive the resonant poles in this case. The methods described here could be considered as an alternative approach for calculation of optical spectra stacked systems.% with arbitrary shifts between the considerable number of relatively simple constituent parts.
% We develop the resonant mode coupling approximation to calculate the optical spectra of a stack of two photonic crystal slabs. The method is based on a derivation of the input and output resonant vectors in each slab in terms of the Fourier modal method in the scattering matrix form. We show that using the resonant mode coupling approximation of the scattering matrices of the upper and lower slabs, one can construct the total scattering matrix of the stack. 
% The formation of the resonant output and input vectors of the stacked system is rigorously derived by means of an effective Hamiltonian. We demonstrate that the proposed procedure dramatically decreases the computation time without sufficient loss of accuracy. We believe that the proposed technique can be a powerful tool for fast solving inverse scattering problems using stochastic optimization methods such as genetic algorithms or machine learning. 
    \end{abstract}        
% \end{frontmatter}

\section{Introduction}
    In the first article in this diptych series \cite{FirstPart}, a resonant mode coupling approximation (RMCA) for the calculation of optical spectra was proposed. {This method appears to be a promising calculation tool because of the great computational speedup of two orders and the in-principle ability to apply it recursively.} Given two subsystems A and B whose scattering matrices could be described in terms of the resonant approximation for the Fourier Modal Method (FMM) \cite{Gippius2010} with slowly varying in energy background scattering matrices $\tilde{\mathrm{S}}^{\mathrm{a,b}}$, resonant energies $\omega_{n}^{\mathrm{a,b}}$ and corresponding output $\left|O_{n}^{\mathrm{a,b}}\right\rangle$ and input $\left\langle I_{n}^{\mathrm{a,b}}\right|$ resonant vectors 
\begin{equation}
\begin{aligned}
&\hspace*{-0.2cm}\left(\begin{array}{c}
\hspace*{-0.13cm}\left|d_{2}\right\rangle \\
\hspace*{-0.13cm}\left|u_{1}\right\rangle
\end{array}\hspace*{-0.13cm}\right)\hspace*{-0.13cm}=\hspace*{-0.13cm}\left[\left(\hspace*{-0.13cm}\begin{array}{cc}
\tilde{\mathrm{S}}_{d d}^{\mathrm{a}} & \tilde{\mathrm{S}}_{du}^{\mathrm{a}} \\
\tilde{\mathrm{S}}_{u d}^{\mathrm{a}} & \tilde{\mathrm{S}}_{u u}^{\mathrm{a}}
\end{array}\hspace*{-0.13cm}\right)\hspace*{-0.13cm}+\hspace*{-0.13cm}\sum_{n=1}^{N}\left|O_{n}^{\mathrm{a}}\right\rangle \frac{1}{\omega-\omega_{n}^{\mathrm{a}}}\left\langle I_{n}^{\mathrm{a}}\right|\right]\hspace*{-0.13cm}\left(\begin{array}{c}
\hspace*{-0.13cm}\left|d_{1}\right\rangle \\
\hspace*{-0.13cm}\left|u_{2}\right\rangle
\end{array}\hspace*{-0.13cm}\right),\\
&\hspace*{-0.2cm}\left(\hspace*{-0.13cm}\begin{array}{c}
\left|d_{3}\right\rangle \\
\left|u_{2}\right\rangle
\end{array}\hspace*{-0.13cm}\right)\hspace*{-0.13cm}=\hspace*{-0.13cm}\left[\left(\hspace*{-0.13cm}\begin{array}{cc}
\tilde{\mathrm{S}}_{d d}^{\mathrm{b}} & \tilde{\mathrm{S}}_{d u}^{\mathrm{b}} \\
\tilde{\mathrm{S}}_{u d}^{\mathrm{b}} & \tilde{\mathrm{S}}_{u u}^{\mathrm{b}}
\end{array}\hspace*{-0.13cm}\right)\hspace*{-0.13cm}+\hspace*{-0.13cm}\sum_{n=1}^{M}\left|O_{n}^{\mathrm{b}}\right\rangle \frac{1}{\omega-\omega_{n}^{\mathrm{b}}}\left\langle I_{n}^{\mathrm{b}}\right|\right]\hspace*{-0.13cm}\left(\hspace*{-0.13cm}\begin{array}{c}
\left|d_{2}\right\rangle \\
\left|u_{3}\right\rangle
\end{array}\hspace*{-0.13cm}\right),
\end{aligned}
\label{resonancesAB}
\end{equation}
one can define the resonant excitation coefficients $\alpha,\beta$ that determine how the incoming electromagnetic waves excite the resonances of the subsystems:

\begin{equation}
\begin{array}{l}
    \alpha_{n}=
    \dfrac{1}{\omega-\omega_{n}^{\mathrm{a}}}\left\langle I_{n}^{\mathrm{a}} \right|
    \begin{pmatrix}
    \left|d_{1}\right\rangle\\
    \left|u_{2}\right\rangle
    \end{pmatrix}\hspace*{-0.1cm}=\dfrac{\hspace*{-0.1cm}\left\langle I_{u, n}^{\mathrm{a}}|u_2\right\rangle\hspace*{-0.07cm}+\hspace*{-0.07cm}\left\langle I_{d, n}^{\mathrm{a}}|d_1\right\rangle}{\omega-\omega_{n}^{\mathrm{a}}},\\
        \beta_{n}=\dfrac{1}{\omega-\omega_{n}^{\mathrm{b}}}
    \left\langle I_{n}^{\mathrm{b}} \right|
    \begin{pmatrix}
    \left|d_{2}\right\rangle\\
    \left|u_{3}\right\rangle
    \end{pmatrix}\hspace*{-0.1cm}=\dfrac{\hspace*{-0.1cm}\left\langle I_{d, n}^{\mathrm{b}}|d_2\right\rangle\hspace*{-0.07cm}+\hspace*{-0.07cm}\left\langle I_{u, n}^{\mathrm{b}}|u_3\right\rangle}{\omega-\omega_{n}^{\mathrm{b}}}.
    \end{array}
    \label{coefs_definition}
\end{equation}
The resonant coefficients should be found from master equation 
\begin{gather}
\label{Eigen_EM}
\omega\left(\begin{array}{c}
\alpha \\
\beta
\end{array}\right)={\underbrace{\left(\begin{array}{cc}
{H}_{aa} &{H}_{ab} \\
{H}_{ba} & {H}_{bb}
\end{array}\right)}_H}
\left(\begin{array}{c}
\alpha \\
\beta
\end{array}\right)+\underbrace{\left\langle I\right|\left(\begin{array}{c}
\left|d_1\right\rangle\\
\left|u_3\right\rangle
\end{array}\right)}_f.
\end{gather}
Here $H$ is an effective Hamiltonian that shows how the resonances of the subsystems hybridize in the presence of interaction
\begin{equation}
    \begin{pmatrix}
    \Omega^{\mathrm{a}}\hspace*{-0.1cm}+\hspace*{-0.1cm}\left\langle I_{u}^{\mathrm{a}}\left|\tilde{\mathrm{S}}_{u d}^{\mathrm{b}}\mathbb{D}_{dd}\right| {O}_{d}^{\mathrm{a}}\right\rangle &
    \hspace*{-0.5cm}\left\langle I_{u}^{\mathrm{a}}| \mathbb{D}_{uu}| {O}_{u}^{\mathrm{b}}\right\rangle\\
    \hspace*{-0.1cm}\left\langle I_{d}^{\mathrm{b}}| \mathbb{D}_{dd}| {O}_{d}^{\mathrm{a}}\right\rangle &
    \hspace*{-0.6cm}\Omega^{\mathrm{b}}\hspace*{-0.1cm}+\hspace*{-0.1cm}\left\langle I_{d}^{\mathrm{b}}\left|\tilde{\mathrm{S}}_{d u}^{\mathrm{a}}\mathbb{D}_{uu}\right| {O}_{u}^{\mathrm{b}}\right\rangle
    \hspace*{-0.07cm}
    \end{pmatrix}
    \hspace*{-0.07cm}\equiv\hspace*{-0.07cm}{\begin{pmatrix}
    \hspace*{-0.07cm} H_{{aa}} & \hspace*{-0.1cm}H_{{ab}}\\
    \hspace*{-0.07cm} H_{{ba}} & \hspace*{-0.1cm}H_{{bb}}
    \hspace*{-0.07cm} \end{pmatrix}},
    \label{Hamiltonian}
\end{equation}
and $\left\langle I\right|$ is related to the excitation of the stacked system by the incoming waves
\begin{equation}
\label{Input}
\left\langle I\right|=
\left(\hspace*{-0.13cm}\begin{array}{cc}
\left\langle I_{d}^{\mathrm{a}}\right|+\left\langle I_{u}^{\mathrm{a}}\right|\tilde{\mathrm{S}}_{ud}^{\mathrm{b}}\mathbb{D}_{dd}\tilde{\mathrm{S}}_{dd}^{\mathrm{a}}& \hspace*{-0.3cm}\left\langle I_{u}^{\mathrm{a}}\right|\mathbb{D}_{uu}\tilde{\mathrm{S}}_{uu}^{\mathrm{b}} \\
\left\langle I_{d}^{\mathrm{b}}\right|\mathbb{D}_{dd}\tilde{\mathrm{S}}_{dd}^{\mathrm{a}} & 
\hspace*{-0.3cm}\left\langle I_{u}^{\mathrm{b}}\right|+\left\langle I_{d}^{\mathrm{b}}\right|\tilde{\mathrm{S}}_{du}^{\mathrm{a}}\mathbb{D}_{uu}\tilde{\mathrm{S}}_{uu}^{\mathrm{b}}
\end{array}\hspace*{-0.13cm}\right).    
\end{equation}

Finally, it was shown that the solution of the scattering problem for the stacked system could be written as

\begin{equation}
\label{GeneralSolution}
\left(\begin{array}{c}
\left|d_{3}\right\rangle \\
\left|u_{1}\right\rangle
\end{array}\right)=\left[\mathrm{S}^{\mathrm{c}}+
\left|O\right\rangle \begin{pmatrix}
    \alpha\\
    \beta
    \end{pmatrix}\right]\left(\begin{array}{c}
\left|d_{1}\right\rangle \\
\left|u_{3}\right\rangle
\end{array}\right).
\end{equation}

Here $\left|O\right\rangle$ represents the contribution of each resonance of the subsystems into the outgoing waves scattered by the stacked system

\begin{equation}
\label{Output}
    \left|O\right\rangle=\left[\begin{pmatrix}
\editR{\mathbb{O}} & \left|O^{\mathrm{b}}_d\right\rangle\\
\left|O^{\mathrm{a}}_u\right\rangle & \editR{\mathbb{O}}
\end{pmatrix} +
\begin{pmatrix}
\tilde{\mathrm{S}}^{\mathrm{b}}_{dd} & \editR{\mathbb{O}}\\
\editR{\mathbb{O}} & \tilde{\mathrm{S}}^{\mathrm{a}}_{uu}
\end{pmatrix} 
\begin{pmatrix}
\editR{\mathbb{I}} & \tilde{\mathrm{S}}_{du}^{\mathrm{a}}\\
\tilde{\mathrm{S}}_{ud}^{\mathrm{b}} & \editR{\mathbb{I}}
\end{pmatrix}
\begin{pmatrix}
\mathbb{D}_{dd} & \editR{\mathbb{O}}\\
\editR{\mathbb{O}} & \mathbb{D}_{uu}
\end{pmatrix}
\begin{pmatrix}
\left|O^{\mathrm{a}}_d\right\rangle & \editR{\mathbb{O}}\\
\editR{\mathbb{O}} & \left|O^{\mathrm{b}}_u\right\rangle
\end{pmatrix}\right],
\end{equation}

and the new background matrix of the stacked system is as follows: 

\begin{equation}
\label{Background}
 \tilde{\mathrm{S}}^{\mathrm{c}}=
    \begin{pmatrix}
    \editR{\mathbb{O}} & \tilde{\mathrm{S}}^{\mathrm{b}}_{du}\\
    \tilde{\mathrm{S}}^{\mathrm{a}}_{ud} & \editR{\mathbb{O}}
    \end{pmatrix}+
    \begin{pmatrix}
\tilde{\mathrm{S}}^{\mathrm{b}}_{dd} & \editR{\mathbb{O}}\\
\editR{\mathbb{O}} & \tilde{\mathrm{S}}^{\mathrm{a}}_{uu}
\end{pmatrix} 
\begin{pmatrix}
\editR{\mathbb{I}} & \tilde{\mathrm{S}}_{du}^{\mathrm{a}}\\
\tilde{\mathrm{S}}_{ud}^{\mathrm{b}} & \editR{\mathbb{I}}
\end{pmatrix}
\begin{pmatrix}
\mathbb{D}_{dd} & \editR{\mathbb{O}}\\
\editR{\mathbb{O}} & \mathbb{D}_{uu}
\end{pmatrix}
\begin{pmatrix}
\tilde{\mathrm{S}}_{dd}^{\mathrm{a}} & \editR{\mathbb{O}}\\
\editR{\mathbb{O}} & \tilde{\mathrm{S}}_{uu}^{\mathrm{b}}.
\end{pmatrix}
\end{equation}

{It is important to highlight, that all previous formulas are written with slowly varying but still energy-dependent background matrices, while resonant vectors $\left|O_{n}^{\mathrm{a,b}}\right\rangle$, $\left\langle I_{n}^{\mathrm{a}}\right|$ of the subsystems characterize particular resonances and thus are constant. We previously solved system \eqref{Eigen_EM} by assuming that the background S-matrices of the subsystems could be considered energy-independent. We calculated their exact values at some background energy $\omega_{bg}$. In that case, the Hamiltonian $H$ and the matrices $\left\langle I\right|$, $\left|O\right\rangle$ are also constant and the solution is straightforward.  Still, the major drawback of the method is calculation inaccuracy that happens due to poor approximation of the background scattering matrix. The exact and approximate optical spectra differ the most on the boundaries of an energy range where the resonant coupling modes approximation is applied, while at the background energy $\omega_{bg}$ the exact and approximate optical properties are identical. Thus, further development of the resonant coupling method should be focused on a better approximation of the background matrix. Two following main ideas are to be further implemented. 

The accuracy of the approximation can be improved by linear approximation of the background matrix in the indicated energy range. This issue could be addressed by taking into account the first-order derivatives of the background matrices. This will give rise to the first-order correction to the Hamiltonian \eqref{Hamiltonian} which should be substituted into equation system \eqref{Eigen_EM}. The background matrix derivatives could be calculated numerically as $\dfrac{\mathrm{S}(\omega_2)-\mathrm{S}(\omega_1)}{\omega_2-\omega_1}$, which means that the new approximation with the first order correction will be exact in two energy points $\omega_1$ and $\omega_2$. 

To further decrease error and make the method applicable to larger energy ranges one has to interpolate the background matrices based on the exact matrices calculated in sample energy points. The number of such sample points still will be much less than the number of query energy points.}

\section{Linear expansion of the effective Hamiltonian}
Suppose we calculated the background scattering matrices $\tilde{\mathrm{S}}^{\mathrm{b,a}}(\omega_{bg})$, $\tilde{\mathrm{S}}^{\mathrm{b,a}}(\omega_{bg}')$ at two energies $\omega_{bg}$ and $\omega_{bg}'$, and $\omega_{bg}'-\omega_{bg}$ is small but finite (we take it to be 5 or 20 meV in the following examples). The choice of right energy step $\omega_{bg}'-\omega_{bg}$ is a debatable question. In that case we can write 
\begin{gather}
\label{LinearExpSbg}
    \tilde{\mathrm{S}}^{\mathrm{b,a}}(\omega)=\tilde{\mathrm{S}}^{\mathrm{b,a}}(\omega_{bg})+ (\omega-\omega_{bg})\partial_\omega\tilde{\mathrm{S}}^{\mathrm{b,a}},\\
    \partial_\omega\tilde{\mathrm{S}}^{\mathrm{b,a}}=\dfrac{\tilde{\mathrm{S}}^{\mathrm{b,a}}(\omega_{bg}')-\tilde{\mathrm{S}}^{\mathrm{b,a}}(\omega_{bg})}{\omega_{bg}'-\omega_{bg}}
\end{gather}
Now that the background matrices are energy-dependent, one can also write the linear energy-dependent part of the effective Hamiltonian \eqref{Hamiltonian}
\begin{equation}
\label{LinearExpH}
    H(\omega)=H_0+(\omega-\omega_{bg})\partial_\omega H, \quad H_0\equiv H(\omega_{bg})
\end{equation}

There are two approaches to further calculate the first-order derivative of the Hamiltonian: to calculate its numerical derivative or to write an expansion of $H$ up to a linear term, neglecting all terms that contain $\omega^2$ and higher orders. It seems like the calculation of numerical derivative is preferable because in that case, one can be sure that the approximation gives exact results at two points, while the expansion of the Hamiltonian up to the linear terms only implies that the result is exact at only one energy point.

Numerical approach to calculation of $\partial_\omega H$ is straightforward: one should calculate $H(\omega_{bg})$ and $H(\omega_{bg}')$ based on the energy-dependent S-matrices \eqref{LinearExpSbg}. As approximation \eqref{LinearExpSbg} is exact at $\omega_{bg}$ and $\omega_{bg}'$, $H(\omega_{bg})$ and $H(\omega_{bg}')$ values are also precise, thus one can supplement equation \eqref{LinearExpH} with 
\begin{equation}
    \partial_\omega H=\dfrac{H(\omega_{bg}')-H(\omega_{bg})}{\omega_{bg}'-\omega_{bg}}.
    \label{dHCalc}
\end{equation}

Now, we return back to energy dependent equation \eqref{Eigen_EM} on the resonant coefficients $Y\equiv(\alpha,\beta)^\mathrm{T}$.
Note that equation \eqref{Eigen_EM} was previously solved with energy-independent $H$, which  included constant background matrices calculated at some background energy $\omega_{bg}$. 

We substitute the linear expansion \eqref{LinearExpH} into \eqref{Eigen_EM} and separate $\omega$-dependent terms on the left side:
\begin{equation}
    [\omega(\mathbb{I}-\partial_\omega H)-(H_0-\omega_{bg}\partial_\omega H)]Y=f(\omega)
\end{equation}
Now one has to multiply the given equation by $(\mathbb{I}-\partial_\omega H)^{-1}$ on the left:
\begin{equation}
    [\omega\mathbb{I}-(\mathbb{I}-\partial_\omega H)^{-1}(H_0- \omega_{bg}\partial_\omega H)]Y%=\\
    =(\mathbb{I}- \partial_\omega H)^{-1}f(\omega)
\end{equation}

One can see that we are dealing with a new renormalized Hamiltonian
\begin{equation}
    \tilde{H}=(\mathbb{I}-\partial_\omega H)^{-1}(H_0- \omega_{bg}\partial_\omega H)
\end{equation}
and a renormalized input $(\mathbb{I}- \partial_\omega H)^{-1}f(\omega)$. We diagonalize the new Hamiltonian 
\begin{equation}
    \tilde{H}\tilde{X}=\tilde{X}\tilde{\Omega}_c
\end{equation}
and write the inverse operator in exactly the same manner as for the energy-independent problem:
\begin{equation}
    [\omega\mathbb{I}-\tilde{H}]^{-1}=\tilde{X}\dfrac{1}{\omega\mathbb{I}-\tilde{\Omega}_c}\tilde{X}^{-1},
\end{equation}
except that the new Hamiltonian and the new right-hand side now reflect a linear correction factors.

The resonant expansion coefficients could be found as
\begin{equation}
    Y=\tilde{X}\dfrac{1}{\omega\mathbb{I}-\tilde{\Omega}_c}\tilde{X}^{-1}(\mathbb{I}-\partial_\omega H)^{-1}f(\omega).
    \label{ResEnergyLinear}
\end{equation}
Here we find that due to the $f(\omega)$ factor the resonant coefficients $Y$ seem to be energy-dependent in a manner that differs from $\dfrac{1}{\omega\mathbb{I}-\tilde{\Omega}_c}$. This should not be because resonant coefficients are defined as \eqref{coefs_definition} and determine the partial contributions of each resonance of subsystems A and B into the
resonances of the stacked system.  This additional energy dependence will be eliminated further as we will see. 
%The scattering problem solution appears in the form: 
% \begin{equation}
%     \left(\begin{array}{c}
% \left|d_{3}\right\rangle \\
% \left|u_{1}\right\rangle
% \end{array}\right)=\tilde{\mathrm{S}}(\omega) \left(\begin{array}{c}
% \left|d_{1}\right\rangle \\
% \left|u_{3}\right\rangle
% \end{array}\right)+\left|O(\omega)\right\rangle Y.
% \end{equation}

Now we are able to write the formula for the scattering matrix using the scattering problem solution of the form \eqref{GeneralSolution}:
\begin{equation}
    \mathrm{S}(\omega)=\tilde{\mathrm{S}}(\omega)+\left|O(\omega)\right\rangle \tilde{X}\dfrac{1}{\omega\mathbb{I}-\tilde{\Omega}_c}\tilde{X}^{-1}(\mathbb{I}-\partial_\omega H)^{-1}\left\langle I(\omega)\right|
\end{equation}
All the energy-dependent terms should be considered linearly-dependent on $\omega$.
According to the previously used method of $\partial_{\omega}H$ calculation one should use the same approach for calculating
\begin{equation}
\begin{gathered}
    \tilde{\mathrm{S}}(\omega)=\tilde{\mathrm{S}}^c(\omega_{bg})+ (\omega-\omega_{bg})\partial_\omega \tilde{\mathrm{S}}^c\\
    \left|O(\omega)\right\rangle=\left|O(\omega_{bg})\right\rangle+(\omega-\omega_{bg})\partial_\omega\left|O\right\rangle\\
    \left\langle I(\omega)\right|=\left\langle I(\omega_{bg})\right|+(\omega-\omega_{bg})\partial_\omega\left\langle I\right|.
    \label{InOutDependent}
\end{gathered}
\end{equation}

Numerical computation of the background S-matrix derivative as well as input and output vectors derivatives fully resembles equation \eqref{dHCalc}, one just should calculate matrices  \eqref{Input},\eqref{Output}, \eqref{Background} at energies $\omega_{bg}$ and $\omega_{bg}'$ using $\tilde{\mathrm{S}}^{\mathrm{b,a}}(\omega_{bg})$ and $\tilde{\mathrm{S}}^{\mathrm{b,a}}(\omega_{bg}')$.

Full equation for the scattering matrix is as follows:
 \begin{multline}
    \mathrm{S}(\omega)=
    %\tilde{\mathrm{S}}(\omega)+\left|O(\omega)\right\rangle \tilde{X}\dfrac{1}{\omega-\tilde{\Omega}_c}\tilde{X}^{-1}(\mathbb{I}-\partial_\omega H)\left\langle I(\omega)\right|=
        \tilde{\mathrm{S}}^c(\omega_{bg})+(\omega-\omega_{bg})\partial_\omega \tilde{\mathrm{S}}^c +\\
        +\left[\left|O(\omega_{bg})\right\rangle
    +(\omega-\omega_{bg})\partial_\omega\left|O\right\rangle\right]
    \tilde{X}\dfrac{1}{\omega\mathbb{I}-\tilde{\Omega}_c}\tilde{X}^{-1}(\mathbb{I}-\partial_\omega H)^{-1}
    \left[\left\langle I(\omega_{bg})\right|+
    (\omega-\omega_{bg})\partial_\omega\left\langle I\right|\right]
    \label{SmatExpanded}
\end{multline}
%\end{widetext}
Obviously, this equation should be transformed so that the linear dependence on energy manifests itself only in the background scattering matrix of the stacked system, exactly as in the resonant expansion of the subsystems' S-matrices. %so that we eventually obtain an approximation of the initial structure, same as for subsystems A and B. 
Moreover, resonant output and input vectors should not depend on $\omega$ as they represent the fundamental properties of the scattering matrix for the particular optical system. In other words, pole {residue}s of a matrix function $\mathrm{S}(\omega)$ should not depend on the parameter $\omega$ or the calculation approach one uses to derive the poles. The desired transformation could be achieved with separation of energy-dependent and energy-independent terms $(\omega-\omega_{bg})\mathbb{I}=(\tilde{\Omega}_c-\omega_{bg}\mathbb{I})+(\omega\mathbb{I}-\tilde{\Omega}_c)$. In what follows we use definitions:
\begin{equation}
 \begin{gathered}
     \delta\left|O^N\right\rangle=\partial_\omega\left|O\right\rangle \tilde{X}\\
     \delta\left\langle I^N\right|=\tilde{X}^{-1}(\mathbb{I}-\partial_\omega H)^{-1}
    \partial_\omega\left\langle I\right|
 \end{gathered}
 \end{equation}
 \begin{equation}
\begin{gathered}
\label{renormOutIn}
    \left|O^N\right\rangle=\left|O(\omega_{bg})\right\rangle \tilde{X} + \delta\left|O^N\right\rangle(\tilde{\Omega}_c-\omega_{bg}\mathbb{I})\\
    \left\langle I^N\right|=\tilde{X}^{-1}(\mathbb{I}-\partial_\omega H)^{-1}
    \left\langle I(\omega_{bg})\right|+
    (\tilde{\Omega}_c-\omega_{bg}\mathbb{I})\delta\left\langle I^N\right|
    \end{gathered}
    \end{equation}
    which allow us to write
  \begin{equation}
\begin{gathered}
    \left|O(\omega)\right\rangle\tilde{X}=\left|O^N\right\rangle+ \delta\left|O^N\right\rangle(\omega\mathbb{I}-\tilde{\Omega}_c)\\
    \tilde{X}^{-1}(\mathbb{I}-\partial_\omega H)^{-1}\left\langle I(\omega)\right|=\left\langle I^N\right|+
    (\omega\mathbb{I}-\tilde{\Omega}_c)\delta\left\langle I^N\right|
    \end{gathered}
    \end{equation}
    
    Substituting this notation into equation \eqref{SmatExpanded} and reducing $\left(\omega_{bg}\mathbb{I}-\tilde{\Omega}_c\right)^{\pm 1}$ terms we also introduce
    \begin{equation}
    \tilde{\mathrm{S}}^N=\tilde{\mathrm{S}}^c+\left|O^N\right\rangle\delta\left\langle I^N\right|
    +\delta\left|O^N\right\rangle\left\langle I^N\right|%+\\
    +\delta\left|O^N\right\rangle \left(\omega_{bg}\mathbb{I}-\tilde{\Omega}_c\right) \delta\left\langle I^N\right|
    \end{equation}
    \begin{equation}
    \partial_{\omega}\tilde{\mathrm{S}}^N=\partial_{\omega}\tilde{\mathrm{S}}^c+\delta\left|O^N\right\rangle  \delta\left\langle I^N\right|
\end{equation}
and write the final resonant approximation for the stacked system with a linear term
\begin{equation}
    \mathrm{S}(\omega)=\tilde{\mathrm{S}}^N+(\omega-\omega_{bg})\partial_{\omega}\tilde{\mathrm{S}}^N+\left|O^N\right\rangle\dfrac{1}{\omega\mathbb{I}-\tilde{\Omega}_c}\left\langle I^N\right|.
    \label{LinExpFin}
\end{equation}
 Here we should underline that the new resonant output and input vectors \eqref{renormOutIn} are directly related to the resonant vectors \eqref{InOutDependent} calculated at the corresponding resonant energies with additional linear dependence of the background matrices taken into account:
 \begin{equation}
 \begin{gathered}
     \left|O^N\right\rangle=\left|O(\tilde{\Omega}_c)\right\rangle \tilde{X}\\
     \left\langle I^N\right|=\tilde{X}^{-1}(\mathbb{I}-\partial_\omega H)^{-1}\left\langle I(\tilde{\Omega}_c)\right|
     \label{InOutInPoles}
 \end{gathered}
 \end{equation}
 This gives us an insight into what we need to calculate to deduce the resonant approximation when the background matrix is represented by an arbitrary function. Once again, the resonant input and output vectors constitute the first-order poles {residue}s of the scattering matrix, that is why they should be calculated at the proximity of the corresponding resonant energies.
 
    \section{Arbitrary approximation of the background S-matrices}
    Let us assume that we have a series of energies $\omega_{bg}^{(i)}$ at which we previously calculated the values of the background matrices $\tilde{\mathrm{S}}^{\mathrm{a,b}}(\omega_{bg}^{(i)})$. Then it is possible to write a simple piecewise-linear approximation for $Re(\omega)\in[\omega^{(i)}_{bg},\omega^{(i+1)}_{bg}]$:
    \begin{equation}
        \tilde{\mathrm{S}}^{\mathrm{a,b}}(\omega)=\tilde{\mathrm{S}}^{\mathrm{a,b}}(\omega_{bg}^{(i)})%+\\
        +\dfrac{\tilde{\mathrm{S}}^{\mathrm{a,b}}(\omega_{bg}^{(i+1)})-\tilde{\mathrm{S}}^{\mathrm{a,b}}(\omega_{bg}^{(i)})}{\omega_{bg}^{(i+1)}-\omega_{bg}^{(i)}}\left(\omega-\omega_{bg}^{(i)}\right)
    \end{equation}
    Although this is the simplest possible approximation, one can choose an arbitrary interpolation function. Still, here we will proceed with this approximation in order to derive two different calculation schemes, one of which is straightforwardly connected to the previously considered approximation with the additional linear term.
    
     We start with the equation on the resonant coefficients with arbitrary $\tilde{\mathrm{S}}^{\mathrm{a,b}}(\omega)$ and $H(\omega)$ functions:
    \begin{equation}
        \begin{pmatrix}
        \alpha\\
        \beta
        \end{pmatrix}\equiv Y=(\omega\mathbb{I}-H(\omega))^{-1}\left\langle I(\omega)\right|\begin{pmatrix}
        |d_1\rangle\\
        |u_3\rangle
        \end{pmatrix}
    \end{equation}
    By utilizing the well-known pole-search procedure \cite{Gippius2010, Bykov2013} one can write a resonant approximation for the inverse operator
    \begin{equation}
        (\omega\mathbb{I}-H(\omega))^{-1}=\tilde{W}(\omega)+\sum_n\dfrac{|O^w_n\rangle\langle I^w_n|}{\omega_n-\omega^c_n}.
    \end{equation}
    The scattering matrix appears as:
\begin{equation}
    \mathrm{S}(\omega)=\tilde{\mathrm{S}}(\omega)+\left|O(\omega)\right\rangle \left(\tilde{W}(\omega)+\sum_n\dfrac{|O^w_n\rangle\langle I^w_n|}{\omega_n-\omega^c_n}\right)\left\langle I(\omega)\right|
\end{equation}

In accordance with \eqref{InOutInPoles} one should calculate the resonant input and output vectors at the corresponding resonant energies, thus:
    \begin{equation}
        {\mathrm{S}}(\omega)=\tilde{\mathrm{S}}^{int}(\omega)
        +\sum_n\dfrac{\left|O(\omega^c_n)\right\rangle|O^w_n\rangle\langle I^w_n|\left\langle I(\omega^c_n)\right|}{\omega_n-\omega^c_n}.
    \end{equation}
    Here the background matrix $\tilde{\mathrm{S}}(\omega)$ is an interpolated function, exact values of which we calculate at a set of background energy points $\omega_{bg}^{(i)}$:
\begin{equation}
    \tilde{\mathrm{S}}^{int}(\omega_{bg}^{(i)})=\tilde{\mathrm{S}}^c(\omega_{bg}^{(i)})%+\\
    +\left|O(\omega_{bg}^{(i)})\right\rangle\left(\omega_{bg}^{(i)}\mathbb{I}-H(\omega_{bg}^{(i)})\right)^{-1}\left\langle I(\omega_{bg}^{(i)})\right|%-\\
    -
    \sum_n\dfrac{\left|O(\omega^c_n)\right\rangle|O^w_n\rangle\langle I^w_n|\left\langle I(\omega^c_n)\right|}{\omega_{bg}^{(i)}-\omega^c_n}
\end{equation}    
Note that 
$$\tilde{\mathrm{S}}^c(\omega_{bg}^{(i)})+\left|O(\omega_{bg}^{(i)})\right\rangle\left(\omega_{bg}^{(i)}\mathbb{I}-H(\omega_{bg}^{(i)})\right)^{-1}\left\langle I(\omega_{bg}^{(i)})\right|$$
is the exact full scattering matrix of the stacked system at $\omega_{bg}^{(i)}$ because we know the exact values of the subsystems' background matrices at these points. Multiplication of the input and output vectors in the form $\left|O(\omega^c_n)\right\rangle|O^w_n\rangle$ and $\langle I^w_n|\left\langle I(\omega^c)\right|$ could be considered as {"second resonant approximation"}. Computation of the resonant expansion of the Hamiltonian $H$ should be much easier than the resonant expansion of scattering matrices as $H$ has a small size equal to the number of resonant poles considered in the problem.

At the same time, one can use the piecewise linear approximation of the background matrices to calculate the resonances of the stacked system according to \eqref{ResEnergyLinear} and \eqref{LinExpFin}, but each resonance and its {residue} should be found independently. First, the resonant energies could be calculated roughly using \eqref{ResEnergyLinear} applied in the center of the calculation region. Then for each energy $\omega_j\in[\omega_{bg}^{(i)},\omega_{bg}^{(i+1)}]$ we apply \eqref{ResEnergyLinear} in this energy region once more to refine the resonant energy value and further calculate the output and input resonant vectors  \eqref{ResEnergyLinear} for this particular resonance. When all resonances are found one can construct a sum of resonant poles. The background matrix could be calculated at energies $\omega_{bg}^{(i)}$, then it should be interpolated.

\section{Numerical example}
\begin{figure}[h]

\begin{minipage}{0.45\linewidth}
  \centering
    \includegraphics[width=1\linewidth]{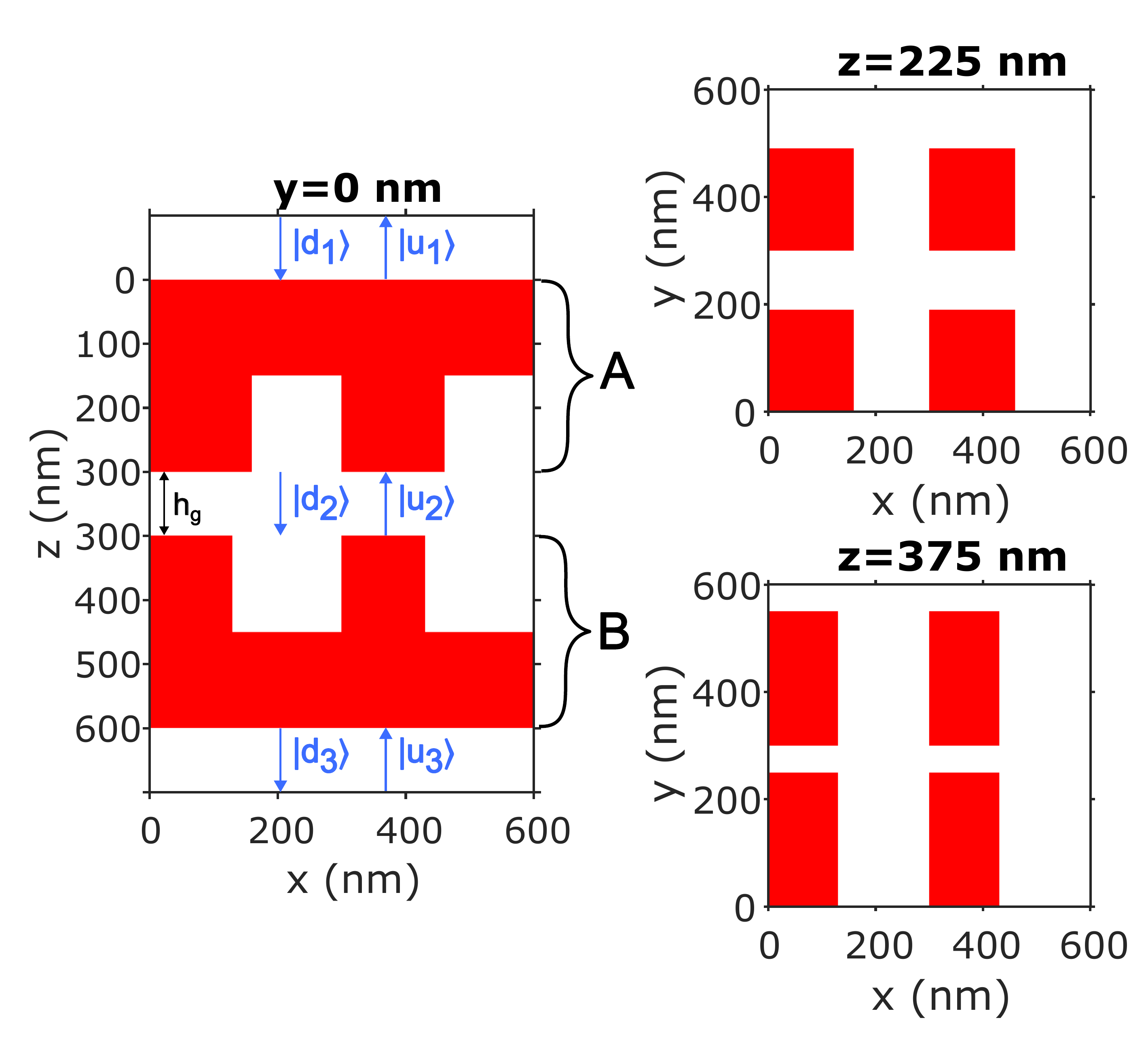}%{Fig1.png}
    \caption{Schematic cross-section of the model structure ($2\times2$ unit cells). Red color denotes the crystalline silicon. The structure is surrounded by the air. The whole structure consists of two subsystems A and B separated by the air gap of the thickness $h_g$ infinitesimally small for maximum interaction of the resonant modes.}% Lower subsystem B is shifted relative to upper system A  at a distance $d_x$, which is 90~nm for the particular configuration.}

    \label{fig:Struct}
\end{minipage}
\hfill
\begin{minipage}{0.53\linewidth}
    \centering
    \includegraphics[width=1\linewidth]{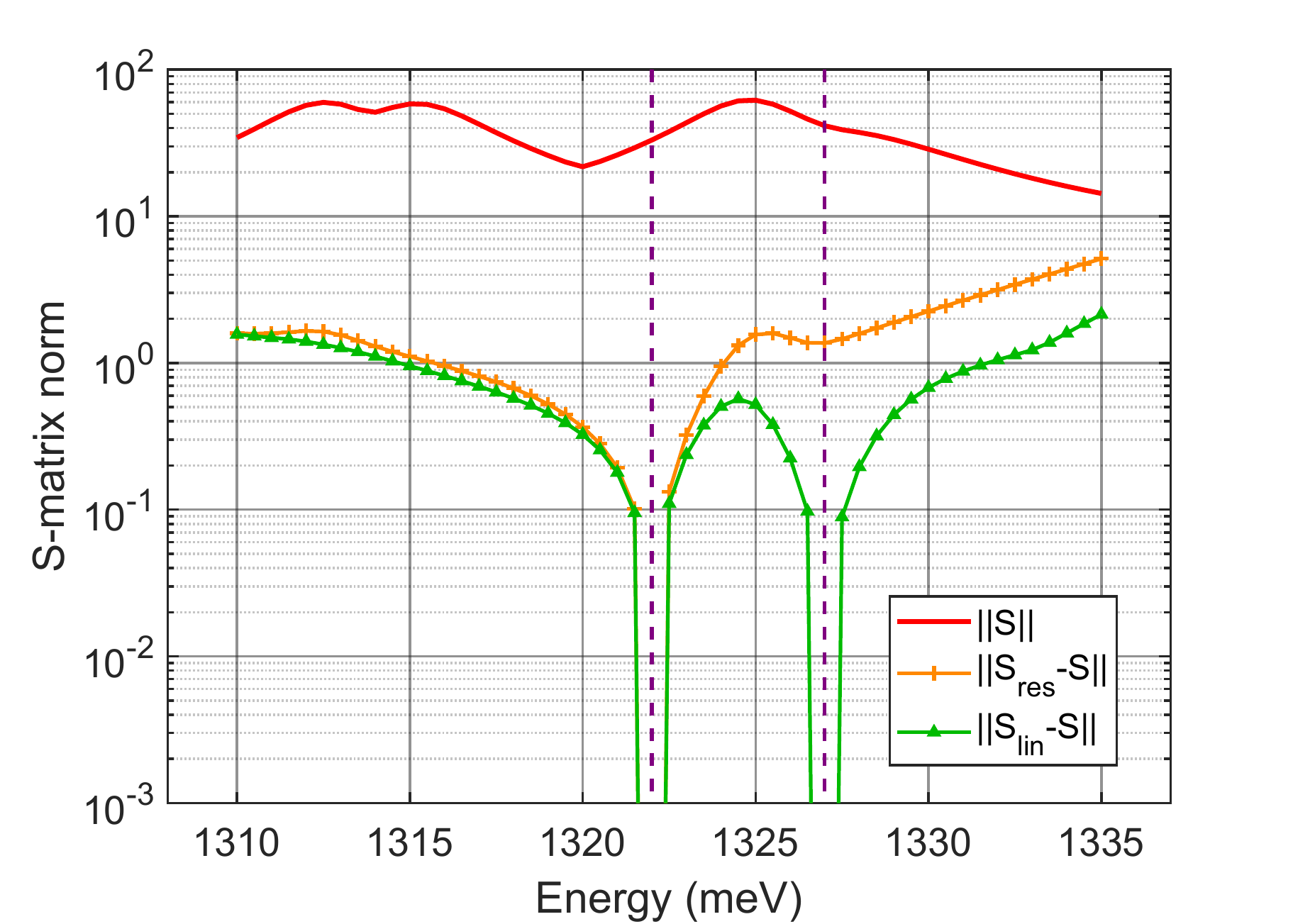}
    \caption{Norm of the exact scattering matrix calculated by the standard
FMM (solid red line) and the 2-norms of the difference between
 the exact scattering matrix and those derived using the resonant mode coupling approximation with energy-independent background (orange crosses) and linear correction term (green triangles). Vertical purple dashed lines denote the background energies $\omega_{bg}$ and $\omega_{bg}'$. The
calculations are conducted for subsystem B located under subsystem A without lateral shift.}
    \label{fig:Spectra}
    \end{minipage}%
\end{figure}

In this section we compare methods efficiencies in terms of approximation precision and time consumption. We choose the structure to be the same as in the first article in this diptych. Two structures A and B consist of 2D periodic gratings located on top of homogeneous waveguiding layers. Pillars of structure A have sizes $160\times190$ nm, while pillars of structure B are $130\times250$ nm wide (see Fig.\ref{fig:Struct}). Both structures have lateral periods equal to $p=300$ nm along lateral directions. All layers are 150-nm thick and made of crystalline Si with permittivity $\varepsilon=13.030+0.033i$. Gratings are directed towards each other, the distance between two subsystems is chosen to be infinitesimally small for maximum modes interaction. The structures are surrounded by air.

Let us first analyze the precision improvements achieved by using a coupled resonant modes approximation with a linear correction to the background matrices instead of the basic approach with energy-independent background matrices. We first take exactly the same problem as in the previous article, which means that we try to retrieve the energy spectra of the stacked system in the energy range $\omega\in[1310,1335]$ meV. The subsystems are not shifted relative to each other; the stacked system is illuminated by plane waves with lateral projections of the wavevector $\{k_x,k_y\}=\dfrac{2\pi}{p}\{0.01,0.02\}$.   One can see in Fig. \ref{fig:Spectra} that with additional linear term the accuracy of the method is be significantly improved in the region $\omega=[1322,1335]$ meV and especially $\omega=[1322,1327]$ meV. Now the scattering matrix is fully retrieved up to machine precision at two two energy points $\omega_{bg}=1322$ meV and $\omega_{bg}'=1327$ meV. 

\begin{figure*}[h]
    \centering
    \includegraphics[width=1\linewidth]{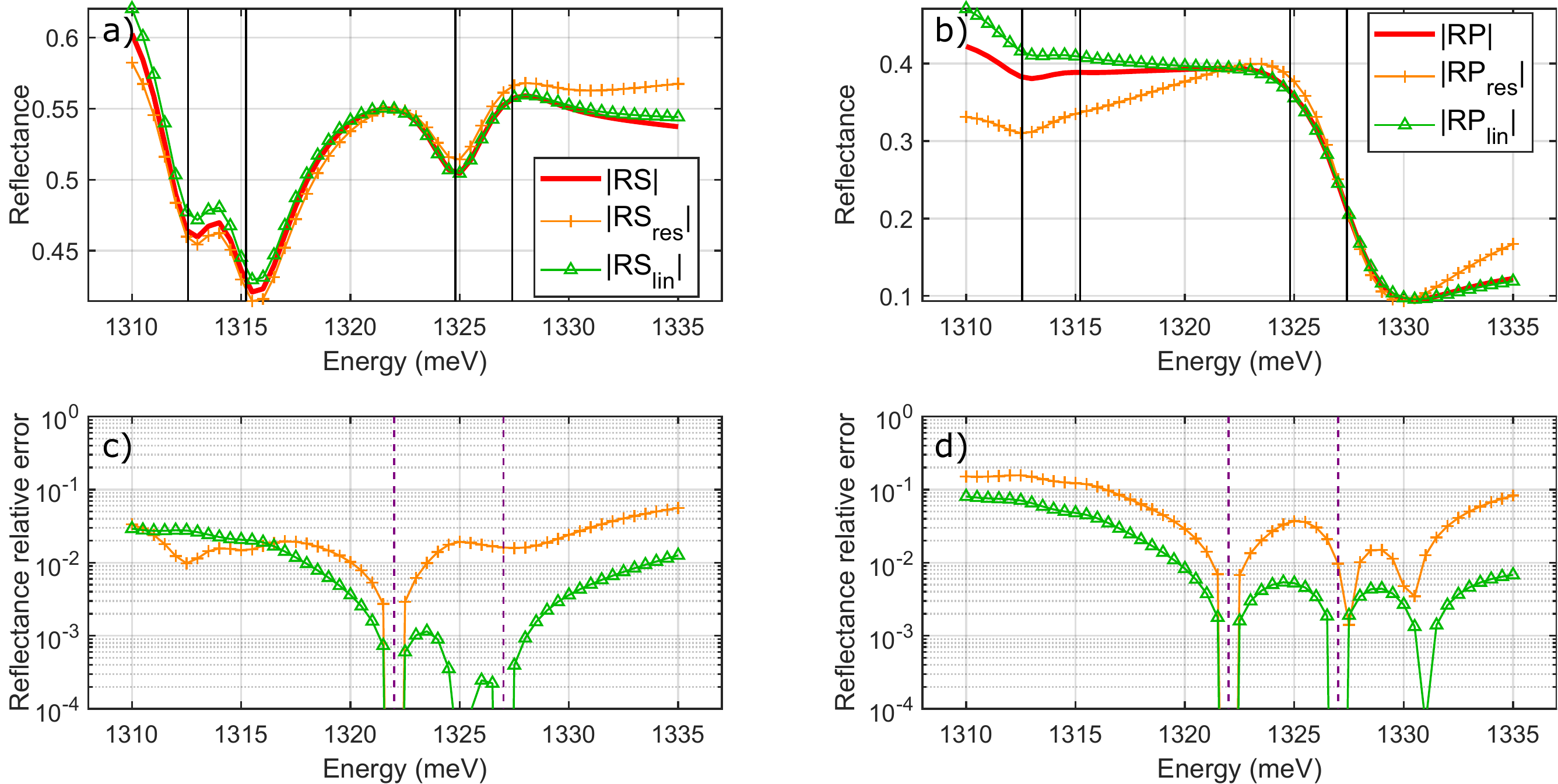}
    \caption{ Panels a) and b) - Reflectance energy spectra in a) S-polarization and b) P-polarization calculated using the
standard FMM  (solid red line) and using the resonant mode coupling approximation with constant background (orange crosses) or with linear term correction (green triangles). Panels c) and d) - modulus of the difference between the exact reflectance and that calculated using the resonant mode coupling approximation with constant background (orange crosses) or with linear term correction (green triangles) in S- and P-polarizations (panels c) and d) respectively).}
    \label{fig:SpectraR}
\end{figure*}

\begin{figure*}[h]
    \centering
    \includegraphics[width=1\linewidth]{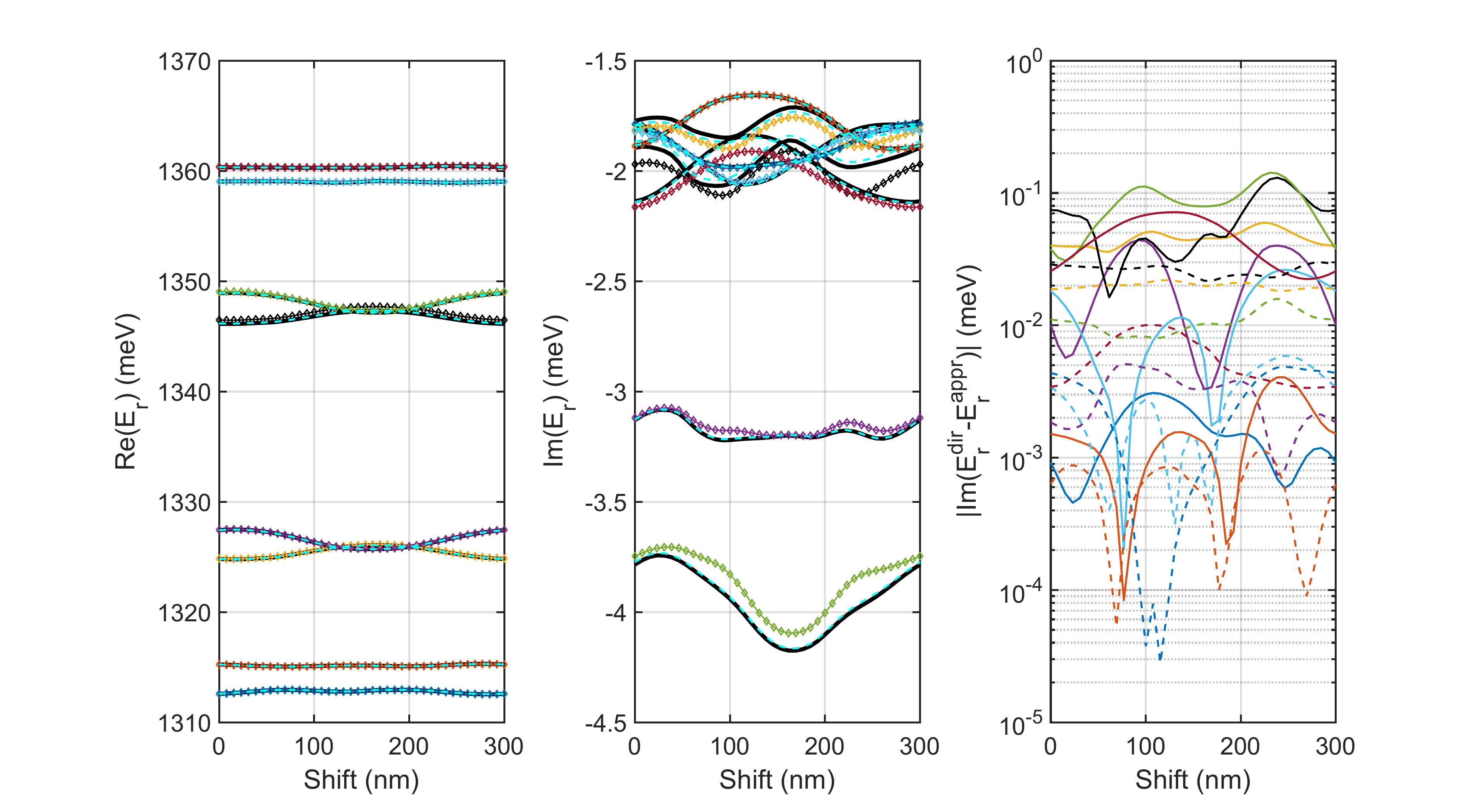}
    \caption{Real (left panel) and imaginary
(central panel) parts of resonant energies of the stacked system
with variable lateral shift $d = d_x = d_y$ of lower subsystem B. Energies are calculated using the standard pole-search procedure (solid black
line) and using the resonant mode coupling approximation with constant background (colored lines with diamonds, $\omega_{bg}=1320$ meV) or with linear term correction (cyan dashed lines, $\omega_{bg}=1320$ meV, $\omega_{bg}'=1340$ meV). Absolute values of difference between the exact and approximate imaginary parts of resonant energies are shown in right panel. Solid lines correspond to constant background approximation, dashed lines correspond to approximation with linear correction. }
    \label{fig:Energies}
\end{figure*}
% \begin{figure*}
%     \centering
%     \includegraphics[width=1\linewidth]{AB_exp_all_NG=11_shift=0_direct.pdf}
%     \caption{}
%     \label{fig:Spectra}
% \end{figure*}

We also compare reflectance spectra in two polarizations: S (electric field perpendicular to the plane of incidence) and P (electric field lies in the plane of incidence) (see Fig. \ref{fig:SpectraR}). The proposed approximation with the energy-linear term in the background matrices is capable to reproduce the exact reflectance coefficient much better than the basic approach does. For example, almost no deviations from the actual reflectance are visible in S-polarization for $\omega>1315$ meV and in P-polarization the method with a linear correction shows consequently smaller error at all energies in the selected range of interest. At the same time, it is obvious that while the linear correction could provide a significant benefit on small energy ranges in comparison with the energy-independent background approximation, evaluation of optical spectra far from the $\omega_{bg}$ will be meaningless due to the following reason. The amplitude reflection coefficient itself is an element of the scattering matrix, thus it is also could be written in the resonant form. As we shift the energy further from the resonance and $\omega_{bg}$ the resonant part becomes weaker, while the term with linear dependence continuously grows. One can always choose energy large enough to make the modulus of reflection coefficient greater than 1, which will be nonphysical and incorrect.

\begin{figure}[h]
    \centering
    \includegraphics[width=0.7\linewidth]{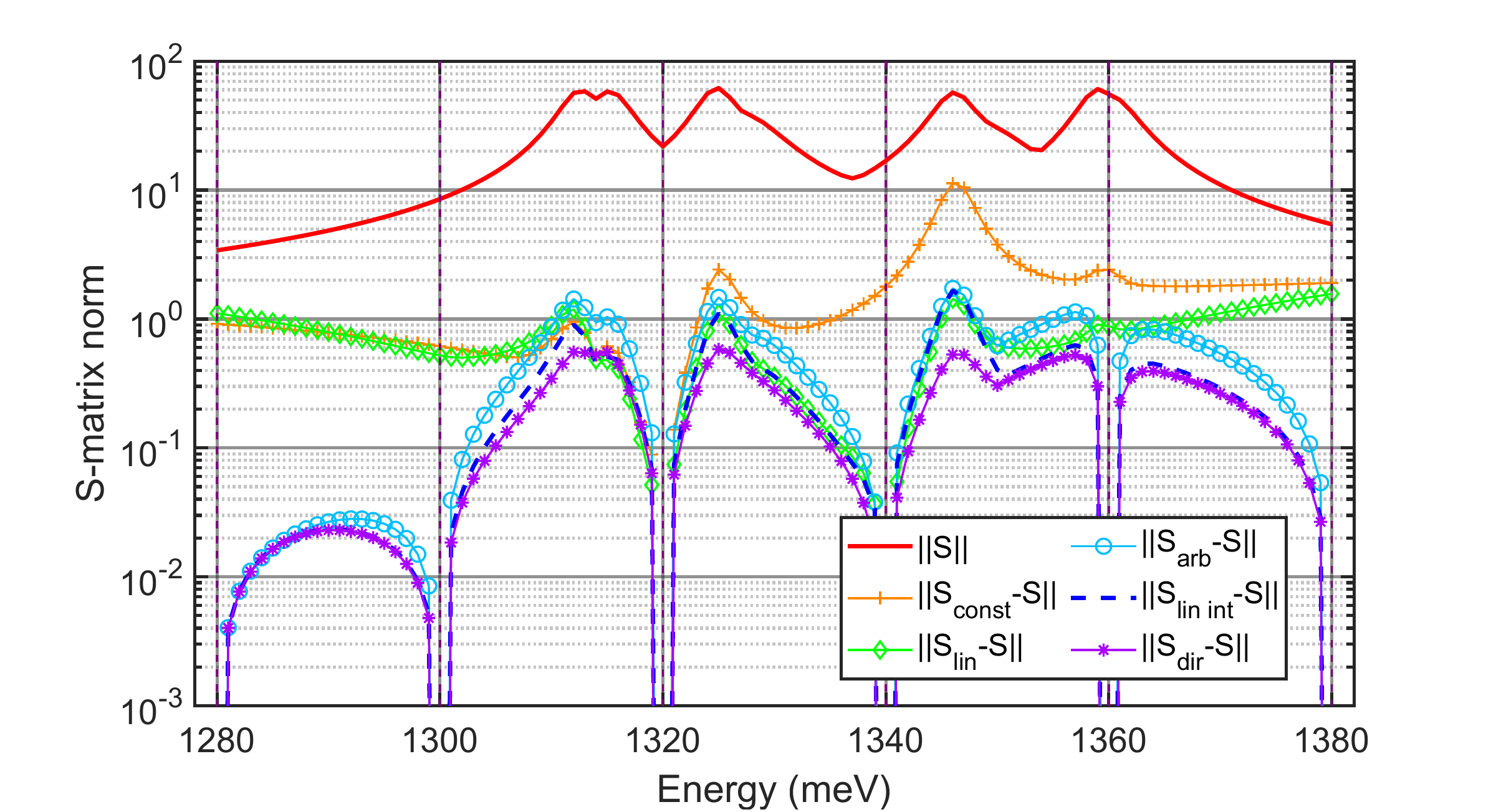}
    \caption{Norm of the exact scattering matrix calculated by the standard
FMM (solid red line); the 2-norms of the difference between
 the exact scattering matrix and those derived using the resonant mode coupling approximation with energy-independent background (orange crosses); with linear correction term (green triangles); piecewise-linear background, resonances found as poles of the effective Hamiltonian (sky-blue circles), or according to \eqref{ResEnergyLinear} and \eqref{LinExpFin} (blue dashed line), direct resonant approximation for the stacked system (magenta stars). Vertical purple dashed lines denote the background energies $\omega_{bg}^{(i)}$. The
calculations are conducted for subsystem B located under subsystem A without lateral shift.}
    \label{fig:SpectraAll}
\end{figure}

\begin{figure}[h]
    \centering
    \includegraphics[width=1\linewidth]{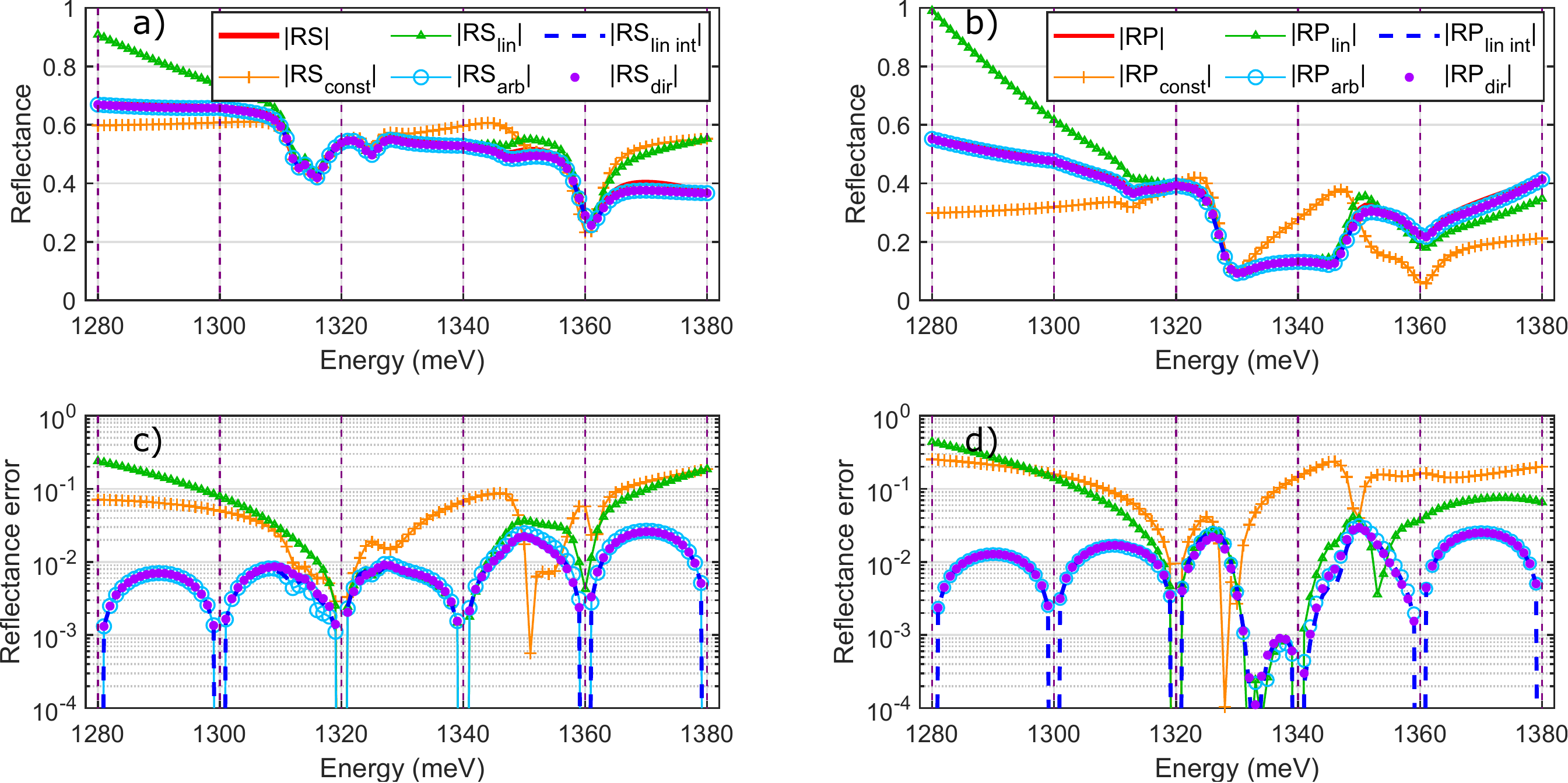}
    \caption{Panels a) and b) - Reflectance energy spectra in a) S-polarization and b)P-polarization calculated using the
standard FMM  (solid red line) and using the resonant mode coupling approximation with different background approximation. Panels c) and d) - modulus of the difference between the exact reflectance and that calculated using the resonant mode coupling approximation in S- and P-polarizations (panels c) and d) respectively). The color legend is exactly the same as in Fig. \ref{fig:SpectraAll}. }
    \label{fig:SpectraAllR}
\end{figure}
To address this issue one may provide either the actual energy-dependence of the background matrices, which is not obvious for complicated periodic structures, or interpolate the background scattering matrices based on the set of scattering matrices calculated in advance for the selected energy range as was proposed in the previous section. Now we take a relatively large energy range $\omega\in[1280,1380]$ meV and calculate the exact background matrices for subsystems A and B at energies $\omega_{bg}^{(i)}=[1280,1300,1320,1340,1360,1380]$ meV. Then the discussed approximate methods are compared to direct resonant expansion of the stacked system which parameters are found through the standard pole-search procedure instead. First, we should note that now we are dealing with 8 resonances in total. The energies of the resonant modes derived with energy-independent background or with linear correction are compared to the exact resonant energies in dependence on the structure B shift. The shifts are performed along both axes of periodicity simultaneously and equally so that $d_x=d_y=d$. One may see that even the energy-independent approach produces the spectral position of the resonances rather precisely, at the same time, the precision of the width calculation of the resonances is far from perfect. The imaginary parts of the resonant energies calculated with the linear correction are closer to the actual ones. The precision of the imaginary part determination could be potentially improved by up to a factor of 10 in some cases. This is especially visible for the wide resonance shown with the green color.

Although the energies of the resonances seem to be found with tolerable accuracy, this does not guarantee that the resonant vectors will be that accurate. One can see in Fig. \ref{fig:SpectraAll} that the resonant modes coupling approximation with a constant background fails to adequately reproduce the resonant peak at 1347 meV - one that should be associated with the green resonance in Fig. \ref{fig:Energies}. The result obtained with additional linear correction does not seem to have this problem. One can even assume that as long the norm of the difference $||\mathrm{S}_{lin}-\mathrm{S}||$ between the approximate and exact scattering matrices does not exceed local maxima in the center of approximation region, the optical spectra will be reproduced with the same accuracy. This is not true, as the difference between the reflectance coefficients is significantly greater at 1280 meV than at 1325 meV (see Fig.\ref{fig:SpectraAllR}), although the norm of the S-matrices difference is approximately the same. We also note that the approximation with constant background is valid only in a small energy range near the background energy $\omega_{bg}=1320$ meV. While it fails to reproduce most of the resonant shapes in the reflectance spectra at 1340-1380 meV, linear correction is capable to highlight all the spectral peculiarities. However, only resonant mode coupling approximation with background matrix interpolation reproduces the whole spectrum. Here it is very important to underline, that the resonant mode coupling approximation accuracy is comparable to the exact resonant approximation calculated for the stacked system from scratch. Moreover, one can be sure that not a single resonant pole is missing as the application of the resonant mode coupling approximation traces even poles with large energy parts, which reveal themselves in broad and smooth energy dependencies.

% \begin{figure*}
%     \centering
%     \includegraphics[width=1\linewidth]{Energies_lin_NG=11_shift=0_direct.pdf}
%     \caption{}
%     \label{fig:Spectra}
% \end{figure*}

\begin{table}[h!]
\centering
\begin{tabularx}{\linewidth}{|X|c|c|} %{|l|X|}
 \hline
 &$N_g=15\times15$ & $N_g=23\times23$ \\
\hline
 $t_{Spectra\, AB}$ - time to calculate $2\times N_E$ S-matrices of subsystems A and B & 317 s & 2808 s\\\hline
 $t_{Find\, Res}$ - time to find 8 resonances of the subsystems with the initial guess values extracted from the shape of peaks in norms of the subsystems' S-matrices & 165 s & 1468 s\\\hline
 $t_{bg}$ -time to calculate the exact background S-matrices at $\omega_{bg}^{(i)}$ & 18.6 s& 166 s\\\hline
 $t_{const}$ - time to calculate parameters of the stacked system resonant approximation + $N_E$ approximate S-matrices and reflection coefficients shown in in Fig. \ref{fig:SpectraAllR} using RMCA with constant background S-matrices & 0.32 + 0.41 s s& 3.2 + 3.5 s\\\hline
 $t_{lin}$ - time to calculate parameters of the stacked system resonant approximation + $N_E$ approximate S-matrices and reflection coefficients shown in in Fig. \ref{fig:SpectraAllR} using RMCA with linear correction to background S-matrices & 1.49 + 0.76 s& 14.3 + 4.1 s\\\hline
 $t_{arb}$ - time to calculate parameters of the stacked system resonant approximation + $N_E$ approximate S-matrices and reflection coefficients shown in in Fig. \ref{fig:SpectraAllR} using RMCA with arbitrary approximation of background S-matrices & 21.5 + 1.9 s& 166.5 + 10.8 s\\\hline
 $t_{lin\,int}$ - time to calculate parameters of the stacked system resonant approximation + $N_E$ approximate S-matrices and reflection coefficients shown in in Fig. \ref{fig:SpectraAllR} using RMCA with piecewise-linear approximation background S-matrices & 7.4 + 1.9 s& 62.2 +10.9 s\\\hline
 $t_{FMM}$ - full time to calculate $N_E$ S-matrices and reflection coefficients of the stacked system using direct FMM for the stacked system & 303 s & 2302 s
% $N_{E}$ & $N_d$ & $t_{FMM}$, s & $t_{Spectra AB}$, s &$t_{Find Res}$, s & $t_{RMCA}$, s\\\hline
% 51& 40 & 5542 & 152 & 93 & 20\\
%  \hline
%  \multicolumn{6}{|c|}{$N_g=23\times23$} \\
% \hline
% $N_{E}$ & $N_d$ & $t_{FMM}$, s & $t_{Spectra AB}$, s &$t_{Find Res}$, s & $t_{RMCA}$, s\\\hline
% 51& 40 & 44840 & 1489 & 719 & 167

\\\hline
\end{tabularx}
\caption{Comparison of the calculation times. Calculations are performed with $N_E=$101 energy points and $N_{bg}=6$ background energy points for methods with interpolation of background S-matrices. $N_g$ is a number of Fourier harmonics used in the calculation.}
\label{time_table}
\end{table}

Finally, we provide Table \ref{time_table} with characteristic computation times. Note that these times are required to calculate only one energy spectra with $N_E=101$ energy points and $N_{bg}=6$ background energy points. At the same time, we propose to use this method only in time-consuming computations when we need to calculate a series of spectral dependencies with varied distances and shifts between the parts of a stacked system or when we can change the number and sequential order of subsystems, as was demonstrated in the first article of this diptych. For such problems acceleration by a factor of 30 is a significant and desired enhancement.

\section{Conclusion}
%%several techniques to improve the accuracy of the resonant mode coupling approximation when approximation of background matrices with energy-independent matrices is insufficient
To sum up, we improved the accuracy of the resonant mode coupling approximation by considering energy-dependent approximation for the background matrices. First, we introduced linear correction to the background scattering matrices and further used it as a base for piecewise-linear and arbitrary approximation of scattering matrices. Even if these two approximations are built on a sparse grid of energy sample points (where the exact background scattering matrices of the subsystems are calculated), they provide a sufficiently small optical spectra calculation error. This allows us to apply the resonant mode coupling approximation in a wide energy range, while the approximation accuracy is controlled by the number of energy sample points. The reasonable density of sample points still remains one or two orders lower than the density of energy points required for good resolution of all spectral peculiarities. Therefore, if the resonant input and output vectors with interpolated background scattering matrices of the subsystems are known, the computation of optical spectra is 30 times faster compared to conventional Fourier Modal Method. 
It is noteworthy that the accuracy of the resonant mode coupling approximation with interpolated background is comparable to the accuracy of the direct resonant approximation for a stacked system. This fact reveals the main advantage of the developed resonant mode coupling approximation that is the possibility to apply this method recurrently, allowing one to describe the multi-stacked system in terms of the subsystems' input and output vectors and background matrices. The formalism of the resonant mode coupling approximation implies that all resonant modes of the stacked system in the energy range of interest are  rigorously tracked and determined. At the same time, the determination of all poles using the standard pole-search procedure becomes more and more complicated as the number of resonances grows with the number of subsystems in a stack. This method paves the way for the new calculation paradigm in which periodic structures are considered as stacks of building blocks --- elementary periodic structures with well-resolved resonances and precalculated accurate resonant approximations. Significant acceleration of computation speed with controllable approximation error and determination of all existing resonances of stacked systems with arbitrary shifts between subsystems makes resonant mode coupling approximation an efficient tool for design optimization. For example, one can prepare a database of optical resonances of elementary photonic structures and then use the resonant mode coupling approximation to calculate the optical spectra of stacked systems. This calculation procedure could be used in machine learning and genetic algorithms to find the best structure design for enhancing a selected optical property.

\section*{Acknowledgments}
{This work is funded by the RSF Grant №22-12-00351.}

\bibliographystyle{elsarticle-num}

\begin{thebibliography}{1}
\expandafter\ifx\csname url\endcsname\relax
  \def\url#1{\texttt{#1}}\fi
\expandafter\ifx\csname urlprefix\endcsname\relax\def\urlprefix{URL }\fi
\expandafter\ifx\csname href\endcsname\relax
  \def\href#1#2{#2} \def\path#1{#1}\fi

\bibitem{FirstPart}
D.~A. Gromyko, S.~A. Dyakov, S.~G. Tikhodeev, N.~A. Gippius
  \href{https://arxiv.org/abs/2110.13647}{Resonant mode coupling approximation for calculation of optical spectra of photonic crystal slabs}, arXiv (2021) 
\newblock \href {https://doi.org/10.48550/arxiv.2110.13647}
  {\path{10.48550/ARXIV.2110.13647}}.
\newline\urlprefix\url{https://arxiv.org/abs/2110.13647}

\bibitem{Gippius2010}
N.~A. Gippius, T.~Weiss, S.~G. Tikhodeev, H.~Giessen,
  \href{http://www.opticsexpress.org/abstract.cfm?URI=oe-18-7-7569}{Resonant
  mode coupling of optical resonances in stacked nanostructures}, Opt. Express
  18~(7) (2010) 7569--7574.
\newblock \href {https://doi.org/10.1364/OE.18.007569}
  {\path{doi:10.1364/OE.18.007569}}.
\newline\urlprefix\url{http://www.opticsexpress.org/abstract.cfm?URI=oe-18-7-7569}

\bibitem{Bykov2013}
D.~A. Bykov, L.~L. Doskolovich,
  \href{http://jlt.osa.org/abstract.cfm?URI=jlt-31-5-793}{Numerical methods for
  calculating poles of the scattering matrix with applications in grating
  theory}, J. Lightwave Technol. 31~(5) (2013) 793--801.
\newline\urlprefix\url{http://jlt.osa.org/abstract.cfm?URI=jlt-31-5-793}

\end{thebibliography}

\end{document}